\documentstyle[12pt]{article}

\newcommand{\beqn}{\begin{equation}}
\newcommand{\eeqn}{\end{equation}}
\newcommand{\beqnarray}{\begin{eqnarray}}
\newcommand{\eeqnarray}{\end{eqnarray}}

\newcommand{\rd}{\partial}
\newcommand{\dfrac}[2]{ \frac{\displaystyle #1}{\displaystyle #2} }

\newcommand{\Res}{\mathop{{\rm Res}}}
\newcommand{\diag}{\mathop{{\rm diag}}}
\newcommand{\Tr}{\mathop{{\rm Tr}}}

\newcommand{\otcomma}{\stackrel{\otimes}{,}}
\newcommand{\bfC}{{\bf C}}
\newcommand{\bfR}{{\bf R}}
\newcommand{\bfZ}{{\bf Z}}
\newcommand{\calM}{{\cal M}}
\newcommand{\calO}{{\cal O}}
\newcommand{\calP}{{\cal P}}
\newcommand{\calT}{{\cal T}}
\newcommand{\Xtilde}{{\tilde X}}
\begin{document}

\title{Gaudin Model, KZ Equation, and \\
Isomonodromic Problem on Torus}
\author{Kanehisa Takasaki\\
{\normalsize Department of Fundamental Sciences, Kyoto University}\\
{\normalsize Yoshida, Sakyo-ku, Kyoto 606, Japan}\\
{\normalsize E-mail: \tt takasaki@yukawa.kyoto-u.ac.jp}}
\date{}
\maketitle
\bigskip

\begin{abstract} 
\noindent
This paper presents a construction of isospectral problems 
on the torus.  The construction starts from an ${\rm SU}(n)$ 
version of the XYZ Gaudin model recently studied by Kuroki 
and Takebe in the context of a twisted WZW model.  In the 
classical limit, the quantum Hamiltonians of the generalized 
Gaudin model turn into classical Hamiltonians with a natural 
$r$-matrix structure.  These Hamiltonians are used to build 
a non-autonomous multi-time Hamiltonian system, which is 
eventually shown to be an isomonodromic problem on the torus. 
This isomonodromic problem can also be reproduced from an 
elliptic analogue of the KZ equation for the twisted WZW model.  
Finally, a geometric interpretation of this isomonodromic problem 
is discussed in the language of a moduli space of meromorphic 
connections. 
\end{abstract}
\bigskip

\begin{flushleft}
KUCP-0111\\
hep-th/9711058
\end{flushleft}
\newpage

\section{Introduction}

It has been argued for the last several years that isomonodromic 
problems are closely related to the Knizhnik-Zamolodchikov 
(KZ) equation \cite{bib:Knizhnik-Zamolodchikov}.  
One of the earliest observations is due to 
Reshetikhin \cite{bib:Reshetikhin}.  Reshetikhin considered 
the Schlesinger equation, and concluded that the KZ equation 
may be viewed as a quantization of the Schlesinger equation.  
Harnad reformulated Reshetikhin's observation in a Heisenberg 
picture \cite{bib:Harnad}.  Let us recall here that the 
Schlesinger equation is an isomonodromic problem on the Riemann 
sphere.  Since the KZ equation on the Riemann sphere can be 
generalized to the Knizhnik-Zamolodchikov-Bernard (KZB) equation 
\cite{bib:Bernard} on the torus, one will naturally expect that 
an associated isomonodromic problem should exist on the torus.  
Korotkin and Samtleben, indeed, derived such an isomonodromic 
problem from the KZB equation of the ${\rm SU}(2)$ WZW model 
\cite{bib:Korotkin-Samtleben}. Recently, Lavin and Olshanetsky 
proposed a general framework for this type of isomonodromic 
problems on a general compact Riemann surface 
\cite{bib:Levin-Olshanetsky}.

We present below an ${\rm SU}(n)$ and ``twisted'' version 
of the isomonodromic problem of Korotkin and Samtleben.  
The word ``twisted'' means that our isomonodromic problem 
is related to the ``twisted WZW model'' recently studied 
by Kuroki and Takebe \cite{bib:Kuroki-Takebe}. 
The method of construction, too, is considerably different.  
Korotkin and Samtleben start from a Hamiltonian formulation 
of the Chern-Simons theory (related to the ordinary untwisted 
WZW model), and formulate the isomonodromic problem in terms 
of a meromorphic connection induced on the torus.  
We follow a more direct approach that has been known for 
a class of isomonodromic problems on the Riemann sphere 
\cite{bib:Jimbo-Miwa,bib:Harnad-dual}.  This class of 
isomonodromic problems can be systematically derived from 
a class of isospectral problems as a non-autonomous analogue; 
of particular interest is the case where the isospectral problems 
are the so called Hitchin systems \cite{bib:Hitchin} and their 
generalizations to punctured Riemann surfaces \cite{bib:Markman}.  
As such an isospectral problem, we now take an ${\rm SU}(n)$ 
version of the XYZ Gaudin model considered by Kuroki and Takebe 
in the study of the twisted WZW model. Like the ordinary Gaudin 
(or ``Calogero-Gaudin'') models 
\cite{bib:FFR,bib:Enriquez-Rubtsov,bib:Nekrasov}, 
this generalized Gaudin model is a quantized Hitchin system 
in the sense of Beilinson and Drinfeld \cite{bib:Beilinson-Drinfeld}. 
In the classical limit, mutually commutative Gaudin Hamiltonians 
turn into Poisson-commutative Hamiltonians with an $r$-matrix 
structure.  We construct from these Hamiltonians a non-autonomous 
multi-time Hamiltonian system, then rewrite it into a Lax system 
of isomonodromic type, and finally confirm that this isomonodromic 
problem is linked, in the sense of Reshetikhin, with Etingof's 
elliptic analogue of the KZ equation \cite{bib:Etingof} 
(the ``elliptic KZ equation'' in the terminology of Kuroki and 
Takebe \cite{bib:Kuroki-Takebe}).  As shown by Kuroki and Takebe, 
this equation plays the role of the KZ equation for the twisted 
WZW model.

Isomonodromic problems on the torus (and more general 
compact Riemann surfaces) have also been studied by 
Okamoto \cite{bib:Okamoto}, Iwasaki \cite{bib:Iwasaki} 
and Kawai \cite{bib:Kawai} by complex analytic and 
geometric methods.  Although their work has been mostly 
focussed on scalar Fuchsian equations, their methods 
can be applied to matrix systems like ours.  
We shall show that this reveals a geometric origin of 
the Hamiltonian structure of our isomonodromic problem.

\section{Generalized Gaudin Model}

We now briefly review Kuroki and Takebe's generalization 
of the XYZ Gaudin model to an ${\rm SU}(n)$ spin system 
\cite{bib:Kuroki-Takebe}.

Let us first introduce basic functions and matrices. 
Let $X$ be the torus (elliptic curve) with modulus $\tau$, 
i.e., $X = \bfC / \bfZ + \bfZ \tau$.   For integer indices 
$(a,b)$, let $\theta_{[ab]}(z)$ denote the special theta functions 
\beqnarray 
    && \theta_{[ab]}(z) = 
    \theta_{\frac{a}{n} - \frac{1}{2},\frac{1}{2} - \frac{b}{n}}(z,\tau), 
        \nonumber \\ 
    && \theta_{\kappa \kappa'}(z,\tau) = 
    \sum_{m\in\bfZ} \exp\bigl[ \pi i \tau (m + \kappa)^2 + 
        2\pi i (m + \kappa) (z + \kappa') \bigr]. 
\eeqnarray
Furthermore, let $J_{ab}$ and $J^{ab}$ be the $n \times n$ matrices 
\beqn
    J_{ab} = g^a h^b, \quad 
    J^{ab} = \frac{1}{n} J_{ab}^{-1}, 
\eeqn
where 
\beqn 
    g = \diag\bigl(1,\omega,\omega^2,\cdots,\omega^{n-1}\bigr), \quad 
    h = \bigl( \delta_{i-1,j}\bigr), \quad 
    \omega = e^{2\pi\sqrt{-1}/n}. 
\eeqn
Note that $g$ and $h$ obey the algebraic relation $gh = \omega hg$. 
It is well known that these $J$'s give a basis of ${\rm su}(n)$ 
(over $\bfR$) and ${\rm sl}(n,\bfC)$ (over $\bfC$). 
$J_{ab}$ and $J^{ab}$ are ``dual'' bases in the sense that 
$\Tr(J_{ab}J^{cd}) = \delta_{ac}\delta_{bd}$.

These functions and matrices are building blocks of 
Belavin's $\bfZ_n$-symmetric $R$-matrix and the associated 
$r$-matrix \cite{bib:Belavin}.  The $R$-matrix is given by 
\beqn
    R(\lambda) = \sum_{(a,b) \in \bfZ_n\times \bfZ_n} 
        W_{ab}(\lambda,\eta) J_{ab} \otimes J^{ab}, 
\eeqn
where 
\beqn
    W_{ab}(\lambda,\eta) = 
        \dfrac{\theta_{[ab]}(\lambda + \eta)}{\theta_{[ab]}(\eta)}. 
\eeqn
These Boltzman weights are $n$-periodic in $a$ and $b$, thereby 
the summation is over $\bfZ_n \times \bfZ_n$. The $r$-matrix is 
the leading nontrivial part in the $\eta$-expansion of $R(\lambda)$ 
at $\eta = 0$: 
\beqn
    r(\lambda) = \sum_{(a,b)\not= (0,0)} 
        w_{ab}(\lambda) J_{ab} \otimes J^{ab}, 
\eeqn
where 
\beqn 
    w_{ab}(\lambda) = \dfrac{\theta_{[ab]}(\lambda) \theta_{[00]}'(0)} 
                      {\theta_{[ab]}(0) \theta_{[00]}(\lambda)}, 
\eeqn
and the prime means $\lambda$-derivative, ${}' = d/d\lambda$.  
The summation in $(a,b)$ is now over 
$\bfZ_n \times \bfZ_n \setminus \{(0,0)\}$.  
We shall frequently omit showing this range explicitly 
(or, equivalently, obey the convention that $w_{00}(\lambda) = 0$). 
The $r$-matrix satisfies the classical Yang-Baxter equation 
\beqn
    \bigl[ r^{(13)}(\lambda), r^{(23)}(\mu) \bigr] = 
    - \bigl[ r^{(12)}(\lambda - \mu), 
         r^{(13)}(\lambda) + r^{(23)}(\mu) \bigr]. 
\eeqn
Here, as usual, the superscript means in which part of the 
tensor product $\bfC^n \otimes \bfC^n \otimes \bfC^n$ the 
$r$-matrix acts nontrivially, e.g., $r^{(12)}(\lambda) 
= \sum w_{ab}(\lambda) J_{ab} \otimes J^{ab} \otimes I$.

The generalized Gaudin model is a limit, as $\eta \to 0$, of 
an inhomogeneous ${\rm SU}(n)$ spin chain with $N$ lattice sites. 
The monodromy matrix of this inhomogeneous spin chain is 
given by 
\beqn
    T(\lambda) = L_N(\lambda - t_N) \cdots L_1(\lambda - t_1), 
\eeqn
where $t_i$'s are inhomogeneity parameters (which eventually 
play the role of time variables in our isomonodromic problem), 
and $L_i$'s are $L$-operators of the form 
\beqn
    L_i(\lambda) = \sum_{(ab)} 
        W_{ab}(\lambda) J_{ab} \otimes \rho_i(J^{ab})
\eeqn
that act on the tensor product $\bfC^n \otimes V_i$ of $\bfC^n$ 
and the representation space $V_i$ of an irreducible representation 
$(\rho_i, V_i)$ of ${\rm su}(n)$.   These $L$-operators satisfy 
the well known equation of ``$RLL = LLR$'' type with $R$ being 
the above $R$-matrix. The monodromy matrix thereby acts on 
$\bfC^n \otimes V$ ($V = V_1 \otimes \cdots \otimes V_N$).  
The leading nontrivial part in the $\eta$-expansion of 
$T(\lambda)$ at $\eta = 0$ can be written 
\beqn
    \calT(\lambda) = \sum_{i=1}^N \sum_{(ab)} 
        w_{ab}(\lambda - t_i) J_{ab} \otimes \rho_i(J^{ab}), 
\eeqn
which satisfies the fundamental commutation relation 
\beqn
    \bigl[ \calT^{(1)}(\lambda), \calT^{(2)}(\mu) \bigr] = 
        - \bigl[ r(\lambda - \mu), 
            \calT^{(1)}(\lambda) + \calT^{(2)}(\mu) \bigr]. 
\eeqn
The superscript now stands for the component of the two 
$\bfC^n$'s in $\bfC^n \otimes \bfC^n \otimes V$ in which 
the monodromy matrix acts nontrivially, e.g., 
$\calT^{(1)}(\lambda) = \sum_i \sum_{(ab)} 
w_{ab}(\lambda - t_i) J_{ab} \otimes I \otimes \rho_i(J^{ab})$.

Mutually commutative Hamiltonians $H_i$ ($i = 0,1,\cdots,N$) 
of the generalized Gaudin model are defined by the relation
\beqn
    \frac{1}{2} \Tr_{n \times n} \calT(\lambda)^2 = 
        \sum_{i=1}^N C_i \calP(\lambda - t_i) + 
        \sum_{i=1}^N H_i \zeta(\lambda - t_i) + H_0, 
\eeqn
where $\calP(z)$ and $\zeta(z)$ are the Weierstrass functions 
with modulus $\tau$.  Note that $C_i$'s are quadratic Casimir 
elements of the algebra generated by $\rho_i(J^{ab})$: 
\beqn
    C_i = \frac{1}{2} \sum_{(ab)} \rho_i(J_{ab}) \rho_i(J^{ab}). 
\eeqn
The $H_i$'s for $i = 1,\cdots,N$ can be written rather explicitly, 
\beqn
    H_i = \sum_{j(\not= i)} \sum_{(ab)} 
        w_{ab}(t_i - t_j) \rho_i(J_{ab}) \rho_j(J^{ab}), 
\eeqn
but explicit expressions of $H_0$ become far complicated. 
There is a linear relation among these Hamiltonians: 
$\sum_{i=1}^N H_i = 0$.

\section{Construction of Isomonodromic Problem on Torus}

\subsection{Classical limit and Poisson bracket} 

We now introduce the following matrix as a classical analogue 
of the $\calT$-operator of the generalized Gaudin model: 
\beqn
    M(\lambda) = \sum_{i=1}^N \sum_{(ab)} 
        w_{ab}(\lambda - t_i) J_{ab} A_i^{ab}. 
\eeqn
$A_i^{ab}$'s are scalar functions of $t_1,\cdots,t_N$ and 
$\tau$.  Our isomonodromic problem will be formulated as 
differential equations for these functions.  (Note that 
$A_i^{ab}$'s depend on $t$ whereas $\rho_i(J^{ab})$'s do not. 
As we shall show later, this amounts to the difference 
of the Heisenberg and Schr\"odinger pictures in quantum 
mechanics.)   We apply the same rule of raising 
and lowering the indices as the $J$-matrices to these 
coefficients.  Thus, e.g., $A_{i,ab}$ are defined by 
$A_{i,ab} = n \omega^{ab} A_i^{-a,-b}$ in accordance 
with the transformation rule for the $J$'s, 
$J_{i,ab} = n \omega^{ab} J^{-a,-b}$.

The commutation relations of $\rho_i(J^{ab})$'s can be 
now translated into Poisson commutation relations among 
the $A$-coefficients.  This Poisson structure can be packed 
into the well known relation 
\beqn
    \{ M(\lambda) \otcomma M(\mu) \} = 
        - \bigl[ r(\lambda - \mu), 
            M(\lambda) \otimes I + I \otimes M(\mu) \bigr]. 
\eeqn
The left hand side, as usual, is an abbreviation of 
$\sum \{M^{ab}(\lambda), M^{cd}(\mu)\} J_{ab} \otimes J_{cd}$, 
where $M^{ab}$ are the coefficients of the expansion 
$M = \sum M^{ab} J_{ab}$.  In terms of the residue matrix 
\beqn
    A_i = \Res_{\lambda = t_i} M(\lambda) 
        = \sum_{(ab)} J_{ab} A_i^{ab}, 
\eeqn
the Poisson structure is nothing but the one induced from 
the Kirillov-Kostant Poisson structure on ${\rm sl}(n,\bfC)$. 
Symplectic leaves of this Poisson structure are the direct 
product $\calO_1 \times \cdots \times \calO_N$ of coadjoint 
orbits $\calO_i$ in ${\rm sl}(n,\bfC)$ on which $A_i$ is living. 
These symplectic leaves become the phase spaces of the 
following non-autonomous Hamiltonian system.

\subsection{Non-autonomous Hamiltonian system} 

Hamiltonians of the classical Gaudin model can be 
obtained as follows: 
\beqn
    \frac{1}{2} \Tr M(\lambda)^2 = 
        \sum_{i=1}^N C_i \calP(\lambda - t_i) + 
        \sum_{i=1}^N H_i \zeta(\lambda - t_i) + H_0. 
\eeqn
$C_i$'s are Casimir elements in the above Poisson algebra. 
The Hamiltonians $H_i$ for $i = 1,\cdots,N$, as in the 
quantum case, can be written explicitly: 
\beqn
    H_i = \sum_{j(\not= i)} \sum_{(ab)} 
        w_{ab}(t_i - t_j) A_{i,ab} A_i^{ab}. 
\eeqn
$H_0$ is also a quadratic form of $A_i^{ab}$'s, though 
its explicit expression is complicated.

It should be noted that $t_i$'s in these formulas are 
just parameters, not time variables, of the classical 
Gaudin system.  The classical Gaudin system is a Hitchin 
system on the punctured torus $X \setminus \{t_1,\cdots,t_N\}$ 
\cite{bib:Enriquez-Rubtsov,bib:Nekrasov}, thus $t_i$'s 
are nothing but the position of punctures.  We now consider 
a non-autonomous multi-time Hamiltonian system with 
the same Hamiltonians, which gives a non-autonomous 
analogue of this Hitchin system.

The non-autonomous Hamiltonian system possesses the 
puncture coordinates $t_1,\cdots,t_N$ and the modulus 
$\tau$ as time variables.  The equations with respect 
to the puncture coordinates are given by 
\beqn
    \dfrac{\rd A_j^{ab}}{\rd t_i} = \{ A_j^{ab}, H_i \}. 
\eeqn
The equation with respect to the modulus, in contrast, 
turns out to take the somewhat strange form 
\beqn
    \dfrac{\rd A_j^{ab}}{\rd \tau} = 
        \Bigl\{ A_j^{ab}, 
           \Bigl(H_0 - \eta_1 \sum_{i=1}^N t_i H_i\Bigr) 
              / 2 \pi \sqrt{-1} \Bigr\} ,  
\eeqn
where $\eta_1$ is the constant (depending only on $\tau$) 
that arises in the transformation law 
\beqn
    \zeta(z + 1) = \zeta(z) + \eta_1 
\eeqn
of the Weierstrass $\zeta$ function.

\subsection{Lax representation} 

In order to confirm that the above non-autonomous 
Hamiltonian system is an isomonodromic problem, 
we now rewrite it into a Lax form.  

A key is the following relation: 
\beqn
    \Bigl\{ M(\lambda), \frac{1}{2}\Tr M(\mu)^2 \Bigr\} = 
    \Bigl[ \sum_{i=1}^N \sum_{(ab)} 
        w_{-a,-b}(\mu - \lambda) w_{ab}(\mu - t_i) J_{ab} A_i^{ab}, 
          M(\mu) \Bigr]. 
\eeqn
This relation can be reduced to a collection of cubic 
relations among $w_{ab}$'s, and can be proven by the same 
function-theoretic method as the proof of the classical 
Yang-Baxter equation (which reduces to quadratic relations 
among $w_{ab}$'s).  

The above formula can be used to rewrite the right hand side 
of the Hamiltonian equations into a matrix commutator.  
First, the residue of both hand sides of the above formula 
at $\mu = t_i$ immediately gives the relation 
\beqn
    \bigl\{M(\lambda), H_i \bigr\} =  
        - \bigl[ A_i(\lambda), M(\lambda) \bigr], 
\eeqn
where 
\beqn
    A_i(\lambda) = \sum_{(ab)} w_{ab}(\lambda - t_i) J_{ab} A_i^{ab}. 
\eeqn
(We have used the reflection property $w_{-a,-b}(-\lambda) = 
- w_{ab}(\lambda)$ \cite{bib:Kuroki-Takebe} of $w_{ab}$'s, too.)
The Poisson bracket with $H_0$ requires lengthy calculations, 
which eventually boil down to the relation 
\beqn
    \{ M(\lambda), H_0 \} = 
        \Bigl[ 4\pi\sqrt{-1} B(\lambda) 
            - \eta_1 \sum_{i=1}^N t_i A_i(\lambda), 
               M(\lambda) \Bigr], 
\eeqn
where 
\beqnarray 
    && B(\lambda) = \sum_{i=1}^N \sum_{(ab)} 
        Z_{ab}(\lambda - t_i) J_{ab} A_i^{(ab)}, 
        \nonumber \\
    && Z_{ab}(\lambda) = \frac{w_{ab}(\lambda)}{4\pi\sqrt{-1}}
        \left( \frac{\theta_{[ab]}'(\lambda)}{\theta_{[ab]}(\lambda)}
              - \frac{\theta_{[ab]}'(0)}{\theta_{[ab]}(0)} \right). 
\eeqnarray

The above formula for $\{M(\lambda),M_0\}$ is interesting in itself 
from two points of view.  Firstly, this formula can be rewritten 
\beqn
    \Bigl\{ M(\lambda), H_0 - \eta_1 \sum_{i=1}^N t_i H_i \Bigr\} = 
        [ 4\pi\sqrt{-1} B(\lambda), M(\lambda) ],  
\eeqn
thus the previous strange linear combination of $H_i$'s 
in the Hamiltonian equation with respect to $\tau$ 
emerges quite naturally.  Secondly, the functions $Z_{ab}$ 
are also basic constituents of the elliptic KZ equation of 
Etingof (see Section 4).

By this formula, the previous non-autonomous Hamiltonian 
system can be rewritten into the following Lax equations: 
\beqnarray
    \frac{\rd M(\lambda)}{\rd t_i} = 
        - [ A_i(\lambda), M(\lambda) ] 
        - \frac{\rd A_i(\lambda)}{\rd \lambda}, 
        \nonumber \\
    \frac{\rd M(\lambda)}{\rd \tau} = 
        [ 2B(\lambda), M(\lambda)] + 
        2 \frac{\rd B(\lambda)}{\rd \lambda}. 
\eeqnarray
The second terms on the right hand side originate in 
differentiating $w_{ab}(\lambda - t_j)$ in $M(\lambda)$ 
by $t_i$ and $\tau$.  The $\tau$-derivatives are converted 
into $\lambda$-derivatives by the heat equation 
\beqn
    \frac{\rd \theta_{[ab]}(\lambda)}{\rd \tau} = 
        \frac{1}{4\pi\sqrt{-1}} 
        \dfrac{\rd^2 \theta_{[ab]}(\lambda)}{\rd \lambda^2}. 
\eeqn

\subsection{Isomonodromic Property} 

As usual, the above Lax equations are integrability 
conditions of a linear system: 
\beqn
    \Bigl( \frac{\rd}{\rd \lambda} - M(\lambda) \Bigr) Y(\lambda) = 0, 
        \quad 
    \Bigl( \frac{\rd}{\rd t_i} + A_i(\lambda) \Bigr) Y(\lambda) = 0, 
        \quad 
    \Bigl( \frac{\rd}{\rd \tau} - 2B(\lambda) \Bigr) Y(\lambda) = 0. 
\eeqn
The first equation is a matrix ODE on the torus with regular 
singular points at $t_1,\cdots,t_N$.  The other equations may 
be interpreted as isomonodromic deformations of this ODE 
in the following sense.

Firstly, the solution of the ODE transforms along a small 
closed path around $\lambda = t_i$ as: 
\beqn
    Y(\lambda) \to Y(\lambda) \Gamma_i. 
\eeqn
$\Gamma_i$ represents the local monodromy around $\lambda = t_i$. 
Similarly, the solution of the ODE transforms along the 
fundamental cycles $\alpha: z \to z + 1$ and 
$\beta:z \to z + \tau$, respectively, as: 
\beqnarray
    Y(\lambda + 1) =  h^{-1} Y(\lambda) \Gamma_\alpha, 
        \nonumber \\
    Y(\lambda + \tau) = g Y(\lambda) \Gamma_\beta. 
\eeqnarray
The extra prefactors $h^{-1}$ and $g$ originate in the 
monodromy of $M(\lambda)$ along those cycles, i.e., 
$M(\lambda + 1) = h^{-1} M(\lambda) h$ and 
$M(\lambda + \tau) = g M(\lambda) g^{-1}$. 
$\Gamma_\alpha$ and $\Gamma_\beta$ represent 
the global monodromy.  ``Isomonodromic deformations''
means that these monodromy data are left invariant 
under deformations.

\section{Relation to Elliptic KZ Equation of Twisted WZW Model}

The elliptic KZ equation of Etingof \cite{bib:Etingof} can be written 
\beqnarray 
    && \Bigl( \kappa \frac{\rd}{\rd t_i} + 
        \sum_{j(\not= i)} \sum_{(ab)} w_{ab}(t_i - t_j) 
            \rho_i(J_{ab}) \rho_j(J^{ab}) 
                \Bigr) F(t_1,\cdots,t_N) = 0, 
                    \nonumber \\
    && \Bigl( \kappa \frac{\rd}{\rd \tau} + 
        \sum_{i,j=1}^N \sum_{(ab)} Z_{ab}(t_i - t_j) 
            \rho_i(J_{ab}) \rho_j(J^{ab}) 
                \Bigr) F(t_1,\cdots,t_N) = 0, 
\eeqnarray
where $\kappa = k + n$, $k$ is the level of the twisted 
WZW model, and the above equation characterizes $N$-point 
conformal blocks with the irreducible representations 
$\rho_1,\cdots,\rho_N$ sitting at the marked points 
$t_1,\cdots,t_N$.

Following Reshetikhin \cite{bib:Reshetikhin}, We now add one more 
marked point at $\lambda$ with the fundamental representation 
$(\bfC^n,{\rm id})$.  The associated $N+1$-point elliptic KZ 
equation becomes 
\beqnarray
    && \Bigl( \kappa \frac{\rd}{\rd \lambda} 
       + \sum_{i=1}^N \sum_{(ab)} w_{ab}(\lambda - t_i) 
            J_{ab} \rho_i(J^{ab}) 
                \Bigr) G(\lambda,t_1,\cdots,t_N) = 0, 
                     \nonumber \\
    &&\Bigl( \kappa \frac{\rd}{\rd t_i} 
       + \sum_{j(\not= i)} \sum_{(ab)} w_{ab}(t_i - t_j) 
            \rho_i(J_{ab}) \rho_j(J^{ab})  
                     \nonumber \\ 
    && \quad 
       + \sum_{(ab)} w_{ab}(t_i - \lambda) 
            \rho_i(J_{ab}) J^{ab} 
                \Bigr) G(\lambda,t_1,\cdots,t_N) = 0, 
                     \nonumber \\
    && \Bigl( \kappa \frac{\rd}{\rd \tau} 
       + \sum_{i,j=1}^N \sum_{(ab)} Z_{ab}(t_i - t_j) 
            \rho_i(J_{ab}) \rho_j(J^{ab}) 
                     \nonumber \\
    && \quad 
       + \sum_{i=1}^N \sum_{(ab)} Z_{ab}(t_i - \lambda) 
            \rho_i(J_{ab}) J^{ab} 
       + \sum_{j=1}^N \sum_{(ab)} Z_{ab}(\lambda - t_j) 
            J_{ab} \rho_j(J^{ab}) 
                     \nonumber \\
    && \quad 
     + \sum_{(ab)} Z_{ab}(0) J_{ab} J^{ab} 
            \Bigr) G(\lambda,t_1,\cdots,t_N) = 0. 
\eeqnarray
Let us now take an invertible ``fundamental solution'' 
$F = \exp S$ \cite{bib:Reshetikhin} of the $N$-point elliptic 
KZ equation, and consider equations for $X = F^{-1}G$.  
This is a kind of ``gauge transformation'' to the $N+1$-point 
elliptic KZ equation, and the outcome is the equations 
\beqnarray
    && \Bigl( \kappa \frac{\rd}{\rd \lambda} 
       + \sum_{i=1}^N \sum_{(ab)} 
           w_{ab}(\lambda - t_i) J_{ab} A_i^{ab} 
             \Bigr) X(\lambda,t_1,\cdots,t_N) = 0, 
                 \nonumber \\
    && \Bigl( \kappa \frac{\rd}{\rd t_i} 
       - \sum_{(ab)} w_{ab}(\lambda - t_i) J_{ab} A_i^{ab} 
             \Bigr) X(\lambda,t_1,\cdots,t_N) = 0, 
                 \nonumber \\
    && \Bigl( \kappa \frac{\rd}{\rd \tau} 
       + 2 \sum_{i=1}^N \sum_{(ab)} 
              Z_{ab}(\lambda - t_i) J_{ab} A_i^{ab} 
          + \sum_{(ab)} Z_{ab}(0) J_{ab} J^{ab} 
              \Bigr) X(\lambda,t_1,\cdots,t_N) = 0, 
\eeqnarray
where 
\beqn
    A_i^{ab} = F(t_1,\cdots,t_N)^{-1} \rho_i(J^{ab}) F(t_1,\cdots,t_N). 
\eeqn
We thus obtain almost the same equation as the isomonodromic 
linear system in the last section, but notice that there are 
a few discrepancies: 
\begin{enumerate}
  \item The last equation contains extra terms (a linear 
  combination of $J_{ab}J^{ab}$). 
  \item The $A_i^{ab}$'s are not scalar functions, but 
  operators on $V = V_1 \otimes \cdots \otimes V_N$. 
  \item The above linear system contains an arbirary parameter $k$. 
  The previous isomonodromic linear system can be reproduced by 
  putting $\kappa = -1$. 
\end{enumerate}

The first discrepancy can be readily remedied, because 
the extra terms are scalar ($J_{ab}J^{ab} = I/n$) and 
can be ``gauded away'' by a scalar gauge transformation 
$X \mapsto e^{f(\tau)} X$ with a function $f(\tau)$ of 
$\tau$ only.  This does not affect the other part of the 
above equations.

The second discrepancy is more essential.  From the 
point of view of Reshetikhin \cite{bib:Reshetikhin} 
and other people \cite{bib:Harnad,bib:Korotkin-Samtleben}, 
the above system is a quantization of the isomonodromic 
problem in the last section.  (This is parallel to the 
relation between quantum and classical Hitchin systems.) 
The operators $A_i^{ab}$ depend on $t$'s and $\tau$, 
but inherit the same commutation relations as the 
$J$-matrices from $\rho_i(J^{ab})$. The passage from 
$\rho_i(J^{ab})$ to $A_i^{ab}$ amounts to the change from 
the Schr\"odinger picture to the Heisenberg picture.  
``Classical limit'' now means replacing these operators by 
functions on a phase space.  This gives our isomonodromic 
problem.

The parameter $\kappa$ plays a role in the limit towards the 
critical level $k = -n$. In this limit, the (classical or quantum) 
isomonodromic problem in the above sense turns into an 
isospectral problem, i.e., a Hitchin system on the punctured 
Riemann surface $X \setminus \{t_1,\cdots,t_N\}$.  A more 
careful analysis of this limit leads to the so called 
``Whitham dynamics'' of the spectral curve, which describes 
the isomonodromic problem as a slowly varying isospectral 
problem \cite{bib:Levin-Olshanetsky,bib:Takasaki}.

\section{Geometric Origin of Hamiltonian Structure}

Our isomonodromic problem may be reformulated in a geometric 
language at least in two different ways.  One option is 
Levin and Olshanetsky's framework \cite{bib:Levin-Olshanetsky} 
based on Hamiltonian reduction. Another option is Iwasaki's 
geometric framework \cite{bib:Iwasaki}, which has been 
successfully applied by Kawai \cite{bib:Kawai} to a 
scalar isomonodromic problem on the torus.  In this section, 
we follow the second approach and show a geometric origin 
of the somewhat involved Hamiltonian structure of our 
isomonodromic problem.

In the geometric approach of Iwasaki and Kawai, 
meromorphic linear ODE's on a compact Riemann surface $X$ 
are converted into meromorphic connections $\nabla$ on a 
holomorphic vector bundle $E$.  All singular points are 
assumed to be regular singular points.  One can then 
develop an analogue of the Kodaira-Spencer theory on 
the moduli space $\calM$ of meromorphic connections with 
a given {\it constant} set of local monodromy data around 
the regular singular points.  The tangent spaces of this 
moduli space, i.e., infinitesimal deformations leaving 
invariant the local monodromy, can be described by the 
language of twisted de Rham cohomologies with coefficients 
in the endomorphism bundle ${\rm End}E$.  A key of Iwasaki's 
ideas is that the Poincar\'e-Lefschetz duality in those 
twisted de Rham cohomologies induces a closed 2-form 
(or a symplectic form) $\Omega$ on the moduli space $\calM$.  
This closed 2-form describes the Hamiltonian 
structure of isomonodromic deformations.  Kawai generalized 
these ideas of Iwasaki to a setting where the Riemann surface 
itself also deforms, and considered, as an example, 
the case of isomonodromic deformations of a second order 
Fuchsian ODE $d^2y / d\lambda^2 = Q(\lambda) y$ on the torus.  
It is this generalization of Iwasaki's framework by Kawai 
that we now attempt to apply to our problem.

It is rather straightforward to reformulate our problem 
into Kawai's framework.  In our case, the endomorphism bundle 
${\rm End}E$ in twisted de Rham cohomologies is replaced by 
the Lie algebra bundle ${\rm sl}(n,\bfC)^{\rm tr}$ of Kuroki 
and Takebe \cite{bib:Kuroki-Takebe}.  This Lie algebra bundle 
is obtained from the trivial bundle ${\rm sl}(n,\bfC) \times \Xtilde$ 
over the universal covering $\Xtilde = \bfC$ of $X$ by identifying 
$(A,\lambda) \sim (hAh^{-1},\lambda + 1) \sim (g^{-1}Ag,\lambda + \tau)$. 
The closed 2-form $\Omega$ is defined by the Poincar\'e-Lefschetz 
pairing $<\delta_1 M, \delta_2 M>$ of two infinitesimal 
deformations $\delta_1 M(\lambda)$ and $\delta_2 M(\lambda)$ 
of $M(\lambda)$ that leave invariant the local monodromy 
around $t_1,\cdots,t_N$: 
\beqn
    \Omega(\delta_1 M, \delta_2 M) = < \delta_1 M, \delta_2 M>. 
\eeqn
Calculation of this pairing, as Kawai illustrated, can be 
reduced to a kind of residue calculus, and eventually 
yields the following expression of $\Omega$: 
\beqn 
    \Omega = 
      \sum_{i=1}^N dH_i \wedge dt_i + 
        d\Bigl(H_0 - \eta_1 \sum_{i=1}^N t_i H_i\Bigr) 
          \wedge \frac{d\tau}{2\pi\sqrt{-1}} 
          - \sum_{i=1}^N \Tr dB_i \wedge dA_i. 
\eeqn
Here $A_i$ is the residue matrix of $M(\lambda)$ at $\lambda = t_i$, 
and $dB_i$ is linked with $dA_i$ as
\beqn
    [A_i, dB_i] = dA_i. 
\eeqn
This result is almost parallel to Kawai's result, but 
the third part on the right hand side does not exist 
in the case of Kawai's isomonodromic problem for 
second order scalar ODE's.  This part is nothing but 
the Kirillov-Kostant symplectic form on the symplectic 
leaves (coadjoint orbits) of ${\rm sl}(n,\bfC)$, 
on which the residue matrix $A_i$ moves as a function 
of time variables.  A similar expression of the fundamental 
2-form for general isomonodromic problems is derived in 
Levin and Olshanetsky's work \cite{bib:Levin-Olshanetsky} 
in the framework of Hamiltonian reduction.

The above expression of $\Omega$ also clearly shows 
the correspondence between the Hamiltonians and 
the time variables.  In particular, one can see 
why we have to take the strange linear combination 
$(H_0 - \sum_{i=1}^N t_i H_i)/2\pi\sqrt{-1}$ of 
$H_i$'s in order to describe the isomonodromic 
deformations with respect to the modulus $\tau$.

\section{Conclusion}

The point of departure of our construction is the 
generalized Gaudin model of Kuroki and Takebe.  
Its classical limit provides a Poisson commutative 
set of Hamiltonians with an $r$-matrix structure.  
The isomonodromic problem is first formulated as 
a non-autonomous multi-time Hamiltonian system, 
then rewritten into a Lax representation.  From 
the Lax representation, we have been able to see 
a link with the elliptic KZ equation of Etingof, 
which advocates Reshetikhin's observation in a 
non-zero genus case.  We have also found a geometric 
interpretation of the Hamiltonian structure of 
this isomonodromic problem in the framework of 
Iwasaki and Kawai.

Our isomonodromic problem is a generalization of 
the Schlesinger equation on the Riemann sphere. 
It should be stressed that constructing such an 
isomonodromic problem on the torus is by no means 
an easy task.  This seems to be one of reasons that 
complex analytic and geometric studies on isomonodromic 
problems have been mostly focussed on scalar equations 
\cite{bib:Okamoto,bib:Iwasaki,bib:Kawai}. The approach 
from the KZ equation is rather suited to dealing with 
matrix equations, and deserves to be pursued further.

An interesting issue in this respect is to construct 
difference and $q$-difference analogues of isomonodromic 
problems from difference \cite{bib:Smirnov} and $q$-difference 
\cite{bib:Frenkel-Reshetikhin,bib:Cherednik} analogues of 
the KZ equation. For instance, Jimbo and Sakai constructed 
a $q$-difference analogue of the sixth Painlev\'e equation 
by a $q$-analogue of the standard method for isomonodromic 
problems \cite{bib:Jimbo-Sakai}.  It seems likely that their 
$q$-difference equation is linked with a $q$-difference 
KZ (qKZ) equation.

The author thanks Hironobu Kimura and Takashi Takebe 
for useful comments and discussion.

\end{document}